\documentclass[article]{aa}
\usepackage{amsmath}
\usepackage{graphicx,txfonts,natbib}
\usepackage{longtable}
\usepackage{lscape}

\begin{document}

\title{Low-amplitude rotational modulation rather than pulsations 
in the CoRoT B-type supergiant
  HD\,46769\thanks{Based on CoRoT space-based photometric data; the CoRoT space
    mission was developed and operated by the French space agency CNES, with
    the participation of ESA's RSSD and Science Programmes, Austria, Belgium,
    Brazil, Germany, and Spain.  Based on observations collected at La Silla
    Observatory, ESO (Chile) with the HARPS spectrograph at the 3.6\,m 
telescope,
    under programme LP185.D-0056.  Based on observations obtained with the
    HERMES spectrograph attached to the 1.2m Mercator telescope, which is
    supported by the Fund for Scientific Research of Flanders (FWO), Belgium,
    the Research Council of KU\,Leuven, Belgium, the Fonds National de la
    Recherche Scientific (FNRS), Belgium, the Royal Observatory of Belgium, the
    Observatoire de Gen\`eve, Switzerland, and the Th\"uringer Landessternwarte
    Tautenburg, Germany. Based on observations obtained with the Narval
    spectropolarimeter at the Observatoire du Pic du Midi (France), which is
    operated by the Institut National des Sciences de l'Univers (INSU).}}

\author{C.~Aerts\inst{1,2}
\and S.~Sim\'on-D\'{\i}az\inst{3,4}
\and C.~Catala\inst{5}
\and C.~Neiner\inst{5}
\and M.~Briquet\inst{6,\thanks{Postdoctoral Fellow, Fonds de la Recherche    
Scientifique (F.R.S.-FNRS), Belgium}}
\and N.~Castro\inst{7}
\and V.~S.~Schmid\inst{1,\thanks{Aspirant PhD Fellow, Fonds voor 
Wetenschappelijk    Onderzoek Vlaanderen (FWO), Belgium}}
\and M.~Scardia\inst{8}
\and M.~Rainer\inst{8}
\and E.~Poretti\inst{8}
\and P.~I.~P\'apics\inst{1}
\and P.~Degroote\inst{1,\thanks{Postdoctoral Fellow, Fonds voor Wetenschappelijk
  Onderzoek Vlaanderen (FWO), Belgium}}
\and S.~Bloemen\inst{1}
\and R.~H.~\O stensen\inst{1} 
\and M.~Auvergne\inst{5}
\and A.~Baglin\inst{5}
\and F.~Baudin\inst{9}
\and E.~Michel\inst{5}
\and R.~Samadi\inst{5}
}

\institute{Instituut voor Sterrenkunde, KU\,Leuven, Celestijnenlaan 200D, 
B-3001
Leuven, Belgium 
\and 
Department of Astrophysics, IMAPP, Radboud University Nijmegen, PO Box 9010,
6500 GL Nijmegen, The Netherlands
\and
Instituto de Astrof\'{\i}sica de Canarias, 38200, La Laguna, Tenerife, Spain
\and
Departamento de Astrof\'{\i}sica, Universidad de La Laguna, 38205, La Laguna,
Tenerife, Spain
\and
LESIA, CNRS UMR8109, 
Universit\'e Pierre et Marie Curie, Universit\'e Denis
Diderot, Observatoire de Paris, 92195 Meudon Cedex, France
\and
Institut d'Astrophysique et de G\'eophysique, Universit\'e de Li\`ege, All\'ee
du 6 Ao\^ut 17 B-4000 Li\`ege, Belgium
\and
Argelander-Institut f\"ur Astronomie der Universit\"at Bonn, 
D-53121 Bonn, Germany
\and
INAF - Osservatorio Astronomico di Brera, via E. Bianchi 46, 23807, Merate, LC,
Italy
\and
Institut d'Astrophysique Spatiale, CNRS/Universit\'e Paris XI UMR 8617, F-091405
Orsay, France
}

\date{Received ; accepted}

\authorrunning{Aerts et al.}
\titlerunning{Low-amplitude rotational modulation in the CoRoT B-type 
  supergiant HD\,46769}

\offprints{conny.aerts@ster.kuleuven.be}

\abstract{} {We aim to detect and interpret photometric and spectroscopic
variability of the bright CoRoT B-type supergiant target HD\,46769
($V=5.79$). We also attempt to detect a magnetic field in the target.}  {We
analyse a 23-day oversampled CoRoT light curve after detrending and
spectroscopic follow-up data using standard Fourier analysis and phase
dispersion minimization methods. We determine the fundamental parameters of
the star, as well as its abundances from the most prominent spectral lines.  
We perform a Monte Carlo analysis of 
spectropolarimetric data to obtain an upper limit of the polar magnetic 
field, assuming a dipole field.}
{In the CoRoT data, we detect a dominant period of 4.84\,d with an
amplitude of 87\,ppm and some of its (sub-)multiples. Given the
shape of the phase-folded light curve and the absence of binary
motion, we interpret the dominant variability in terms of rotational
modulation, with a rotation period of 9.69\,d.  
Subtraction of the rotational modulation signal does not
reveal any sign of pulsations. Our results are consistent with the
absence of variability in the Hipparcos light curve.  The
spectroscopy leads to a projected rotational velocity of
72$\pm 2$\,km\,s$^{-1}$ and does not reveal periodic variability or the
need to invoke macroturbulent line broadening. 
No signature of a magnetic field is detected 
in our data. A field stronger than  $\sim 500$\,G at the poles 
can be excluded, unless the 
possible non-detected field were more complex than dipolar.}  {The absence of
pulsations and macroturbulence of this evolved B-type supergiant
is placed into the context of instability computations and of observed
variability of evolved B-type stars.}


\maketitle
%

\section{Introduction}

High-precision space asteroseismic data delivered by the MOST (Walker et
al.\ 2003), CoRoT (Auvergne et al.\ 2009), and {\it Kepler\/} (Gilliland et
al.\ 2010) missions have revolutionised our capacity to probe the interior
physics and evolutionary status of various types of stars. For a few selected
highlights concerning main-sequence stars with masses ranging from solar to
25\,M$_\odot$, as well as low-mass evolved stars in various nuclear burning
stages, see, e.g.\ Michel et al.\ (2008), Huat et al.\ (2009), Degroote et
al.\ (2010), Briquet et al.\ (2011), Bedding et al.\ (2011), Charpinet et
al.\ (2011), Michel \& Baglin (2012), 
Beck et al.\ (2012), Chaplin \& Miglio (2013).
Despite the immense and fast progress, the evolutionary models for some types of
stars remain uncalibrated in terms of high-precision asteroseismic
constraints. This is particularly the case for evolved massive stars, while
models of such objects have a large impact on various research areas of
astrophysics, such as the computation of models with time-variable mass loss for
core-collapse supernovae progenitors or the estimation of the chemical
enrichment of galaxies.

Only three massive supergiants have been studied in detail from uninterrupted,
time-resolved high-precision space photometry so far. The first case concerns
the B2Ib/II star HD\,163899, for which Saio et al.\ (2006) reported the
discovery of tens of pressure and gravity modes with frequencies below
2.8\,d$^{-1}$ (32.4\,$\mu$Hz), and amplitudes below a few mmag from a 37\,d MOST
light curve. Although hot supergiant models for masses between 20 and
24\,M$_\odot$ and their pulsation predictions showed qualitative agreement with
the observations, the interior physics of those models could not be improved due
to the lack of high-precision spectroscopy. Such spectroscopy is 
necessary for good fundamental
parameter and abundance estimation as well as for mode identification of the
detected pulsation frequencies. The second supergiant monitored from space is
the B6Ia star HD\,50064, which was observed by CoRoT for 137\,d and followed up
with high-resolution spectroscopy covering a time base of 169\,d (Aerts et
al.\ 2010b). This supergiant was found to have a variable-amplitude monoperiodic
radial strange-mode pulsation with a period of 37\,d, which is connected with
the variable mass loss of the star. While its 
monoperiodic character did not allow
asteroseismic tuning of its interior structure, the interpretation of the
observations was recently confirmed by evolutionary model computations for
He-core burning models re-entering a blue loop in the Hertzsprung-Russell (HR)
diagram after having passed the red-supergiant phase (Saio et al.\ 2013).
In the third case of HD\,34085 (Rigel, B8Ia), 
the MOST light curve of 28\,d was too short
to uncover the oscillations found in long-term spectroscopy
(Moravveji et al.\ 2012).

Clearly, there is a need to study supergiant pulsations in much more detail to
fine-tune evolutionary models in the upper HR\,diagram. The {\it Kepler\/}
mission cannot help in this respect as there are no hot massive supergiants in
its fixed field of view. It is in this context that we studied the bright
massive B8Ib supergiant HD\,46769 as secondary asteroseismology target with
CoRoT. The fundamental parameters of this star were previously derived by
Lefever et al.\ (2007), based on high-precision spectroscopy containing H and He
lines only. They obtained $T_{\rm eff}=12\,000\,$K, $\log\,g=2.55$, $v\sin
i=68\pm 5\,$km\,s$^{-1}$, and $\log (L/L_\odot)=4.22$.  These authors also
examined the Hipparcos data and did not find any significant variability.
Zorec et al.\ (2009), on
the other hand, applied the Barbier-Chalonge-Divan spectrophotometric
classification method and found the effective temperature to be some 2\,000\,K
higher. They assigned
a spectral type of B7III/II to the star without providing an
estimate for $\log\,g$. In any case, irrespective of the precise effective
temperature and luminosity class, the star is situated outside any known
instability strip. This was also the case for HD\,163899, which was
originally measured by MOST as a standard star. Thus, in view of the lack of
supergiant asteroseismology so far and because it was the only hot massive
supergiant in the pointed fields of CoRoT, HD\,46769 was selected as secondary
target for the mission.

\section{CoRoT light curve analysis}

The supergiant
HD\,46769 was observed by CoRoT during a short run in the anticentre direction
of the Milky Way (SRa03) in March 2010.  The asteroseismology runs deliver data
with a cadence of 32\,s, and SRa03 lasted for 26.4\,d, from HJD\,2455258.426769
until HJD\,2455284.802757. A strong discontinuity connected with instrumental
effects (Auvergne et al.\ 2009) occurs in the light curve near day HJD\,2455262,
after which both the mean flux level and the nature of the trend in the flux
changed.  As such, we restricted our analysis to the last 22.7\,d of the light
curve, which contains 53927 data points of good quality and has a Rayleigh limit
of $1/\triangle T = 0.044\,$d$^{-1}$.  The raw light curve shows an instrumental
downward trend as present in most of the CoRoT light curves and interpreted as
CCD ageing (Auvergne et al.\ 2009). Given that intrinsic variability occurs with
a time scale similar to 
the duration of the data, which we do not want to
affect, we only removed a linear trend before starting the frequency
analysis. For the latter, we adopted the usual zero-point for the time stamps
applied by the CoRoT consortium, i.e.\  1 January 2000 at 12h UT, which
corresponds with HJD$_0$=2451545.0.
\begin{figure}[t]
\begin{center}
\rotatebox{270}{\resizebox{6cm}{!}{\includegraphics{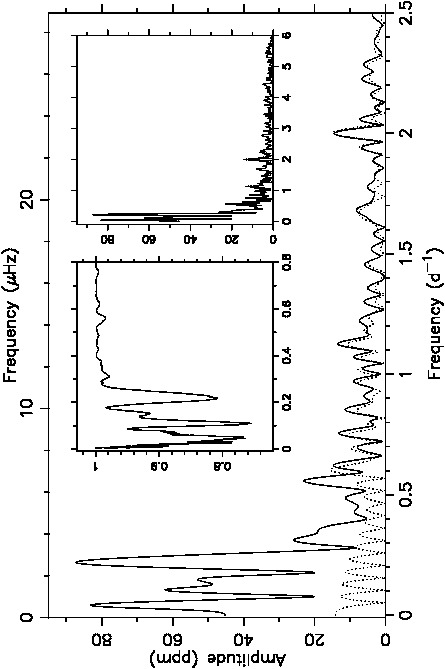}}}
\end{center}
\caption{Outcome of the frequency analysis for the CoRoT light curve of
  HD\,46769. The main panel shows the Scargle periodogram over
  $[0,2.5]\,$d$^{-1}$ (full line) and the version after prewhitening with
  $f=0.0516\,$d$^{-1}$ and seven of its harmonics as 
listed in Table\,\protect\ref{sn} (dotted line).
  The left inset shows the $\Theta\,$statistic and the right inset shows the
  Scargle periodogram over $[0,6]\,$d$^{-1}$. No significant power occurs at
  frequencies above 0.6\,d$^{-1}$, unless they are of instrumental origin (such
  as the peak near 2\,d$^{-1}$).  }
\label{ft}
\end{figure}

\begin{figure*}[t]
\begin{center}
\rotatebox{270}{\resizebox{13.cm}{!}{\includegraphics{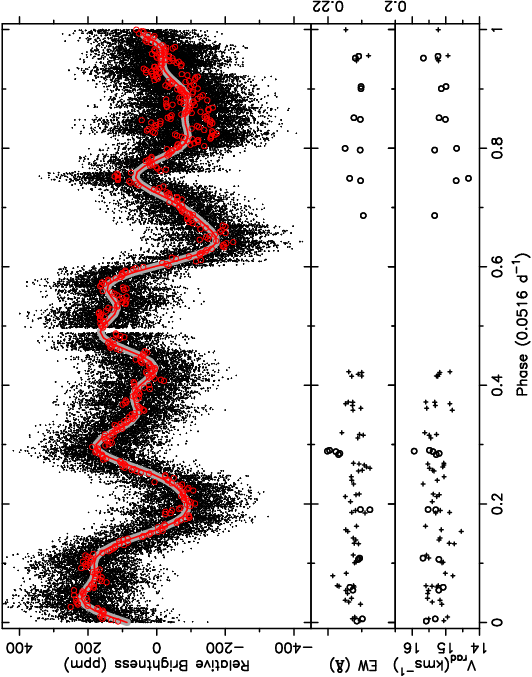}}}
\end{center}
\caption{Upper: the linearly detrended CoRoT light curve of HD\,46769 phased
  with respect to the frequency $f=0.0516\,$d$^{-1}$.  The grey thick full line
  is a harmonic fit with $f$ and seven of its harmonics, of which
  $4f=0.2064\,$d$^{-1}$ is the dominant frequency in the light curve.  The red
  symbols and red line denote averages over 101 consecutive data points and a
  fit to that binned light curve respectively.  Middle and bottom: the
  equivalent width and radial velocity of the \ion{Mg}{ii}\,4481\AA\ line
  as observed in the HERMES (circles) and HARPS (plus signs) spectra.  }
\label{lc}
\end{figure*}

We computed both the Scargle periodogram (Scargle 1982) and the
$\Theta\,$statistic in the framework of phase dispersion minimization (PDM,
Stellingwerf 1978).  Maximum power in the Scargle periodogram occurs at the
frequency near 0.21\,d$^{-1}$, while $\Theta$ reaches a minimum near
0.11\,d$^{-1}$ (Fig.\,\ref{ft}).  Both the Scargle and PDM periodograms also
have a peak near 0.05\,d$^{-1}$, corresponding with a periodicity of 19.4\,d.
The Scargle frequency at maximum amplitude is equal to four times the one of the
second highest amplitude, within the frequency resolution.

In the next section, we add the spectroscopic time series to the photometry to
end up with the most likely frequency $4f$, with value $f=0.0516\,$d$^{-1}$.
We thus made a sinusoidal fit with $4f$ to the light curve. Subsequent
prewhitening leads, within the frequency resolution, to $f$, $2f$, $5f$, $3f$,
and $11f$ respectively, reflecting the strongly non-sinusoidal shape of the
light curve (Fig.\,\ref{lc}).  The frequencies in the data with a significance
above 4\,$\sigma$ are listed in Table\,\ref{sn}. After prewhitening with those
frequencies, we find a residual frequency at
2\,d$^{-1}$.  This frequency, which is well visible in
Fig.\,\ref{ft}, is a known artefact related to day-versus-night variations
connected with CoRoT's low-Earth orbit (Auvergne et al.\ 2009). The satellite
orbital frequency itself is 13.97\,d$^{-1}$, and its harmonics also occur in
CoRoT light curves. However, 
this regime in frequency is not relevant for our study
of HD\,46769.

\begin{table}
  \caption{\label{sn}The frequencies of HD\,46769 
    in the CoRoT light curve along with their amplitudes and
    significance level.} 
\begin{center}
\begin{tabular}{ccc}
\hline\hline
Frequency & Amplitude & Significance \\
(d$^{-1}$) & (ppm) & ($\times$ noise level) \\
\hline
$f=0.0516$ & 67.3$\pm 0.6$ & 14.0 \\
$2f$ &  63.2$\pm 0.6$ & 13.2 \\
$3f$ & 42.9$\pm 0.6$ & 8.9 \\
$4f$ & 86.9$\pm 0.7$ & 18.1 \\
$5f$ & 44.9$\pm 0.7$ & 9.4 \\
$6f$ & 23.2$\pm 0.6$ & 4.8 \\
$7f$ & 21.4$\pm 0.8$ & 4.5 \\
$11f$ & 25.1$\pm 0.7$ & 5.2\\
\hline
\end{tabular}
\end{center}
\tablefoot{The significance level is 
    expressed in terms of the average noise level in the
    Scargle periodogram, which was computed as the average amplitude in the
    range $[0,2]\,$d$^{-1}$ after prewhitening of all the frequencies listed
    here and amounts to 4.8\,ppm.}
\end{table}

Given that the data is oversampled compared with the dominant periodicities, we
repeated the frequency analysis after binning the light curve over 101,
respectively 51, 
data points.  For both cases, these binned light curves gave the
same frequency results as those listed in Table\,\ref{sn} within the
errors. The 101--binned phased light curve and its fit are shown in red in
Fig.\,\ref{lc}.

We subsequently re-analysed the Hipparcos data (Perryman 1997)
of the star and did not find any
significant variability. This is in agreement with the results by Lefever et
al.\ (2007), who used HD\,46769 as a constant comparison star in their study of
28 periodically variable B supergiants, whose variability they interpreted as
being caused by gravity-mode pulsations. The highest peak in the Hipparcos
Scargle periodogram for HD\,46769 occurs near 9\,d$^{-1}$; however it reaches an
amplitude of 4.6\,mmag and is not significant in that data set. We explicitly
checked that it does not occur in the CoRoT light curve.  Given that the
Hipparcos data can only reveal variability at a level of mmag (i.e.\  ppt) or
higher, it is not surprising that we cannot detect the dominant frequency found
in the CoRoT light curve in that data set.

We conclude that the CoRoT light curve of HD\,46769 reveals it to be a
monoperiodic variable with a period of 4.84$\pm0.2$\,d having an amplitude of
87\,ppm and that (sub-)multiples of this period also occur, representing
non-sinusoidal variability undetectable by ground-based photometry. The shape of
the light curve, with deep and less deep minima that re-occur, leads us to
suggest that $2f$ corresponds with the rotation frequency of the star. This
translates into a rotation period of 9.69\,d, although we cannot firmly exclude
$4f$ to be the rotation frequency. The stellar parameters and $v\sin i$ derived
by Lefever et al.\ (2007) do exclude $f$ to be the rotation frequency.


\section{Ground-based follow-up data}

In order to reach a unique interpretation of the variability found in the
CoRoT light curve, with rotation and binarity as most likely scenarios from the
CoRoT photometry, we set up a follow-up spectroscopic campaign. We first made
use of the HERMES spectrograph (Raskin et al.\ 2011) attached to the 1.2m
Mercator telescope at La Palma and assembled 24 spectra spread over one year, in
the period February 2010 until January 2011. The spectra have a high resolving
power 
of $85\,000$ and a typical average signal-to-noise ratio (S/N)  between 100
and 200 in the wavelength range $[4000,5000]\AA$.  At first sight, these data
excluded binarity and did not reveal any long-term periodic spectroscopic
variability. To enhance the detection capabilities and search for short-period
variability as found in the CoRoT light curve, the star was subsequently
included in the HARPS Large Programme for the ground-based preparatory and
follow-up observations for the CoRoT space mission for more intensive monitoring
of several spectra per night during a 25-day campaign. This led to an additional
data set of 62 spectra taken in the HARPS HAM mode (resolving power of 
115\,000), assembled in
December 2012 -- January 2013. These spectra are of similar S/N as
the HERMES spectra.  For both campaigns, the exposure time was of the order
of a few minutes and was adapted to the atmospheric conditions.  We refer to
Table\,\ref{logbook} for a summary of the spectroscopic observations.

The pipeline-extracted spectral orders of the HARPS measurements were merged and
treated together with the pipeline-reduced HERMES exposures. All spectra were
normalised following the procedures developed and outlined in P\'apics et
al.\ (2011, 2012, 2013). The HERMES and HARPS spectra were found to be fully
compatible with each other and could be merged into a single time series data
set with a total time span of 1057\,d, leading to a Rayleigh limit of
0.00095\,d$^{-1}$. From these data, we can 
exclude that HD\,46769 is a binary with an
orbital period below several thousands of days.

\begin{table*}
\caption{\label{logbook}Logbook of the spectroscopy gathered for HD\,46769.}
\begin{center}
\tabcolsep=4pt
\begin{tabular}{llclllrl}
\hline\hline
Telescope & Instrument & Wavelength Range & HJD start & HJD end & N &
$\lambda/\Delta\lambda$  & S/N range  \\ 
\hline
1.2m Mercator, La Palma & HERMES & $[3700,9000]$\,\AA & 5240.472 & 5581.636 & 24 & 85\,000 & 100 -- 200  \\
3.6m ESO, La Silla & HARPS & $[3700,6800]$\,\AA & 6272.664 & 6297.825 & 62 &
115\,000 & 120 -- 250   \\
\hline
\end{tabular}
\end{center}
\tablefoot{The HJD was subtracted with 2450000, N denotes the number of spectra,
  $\lambda/\Delta\lambda$ stands for the resolving power; the listed range in
  S/N was computed in the wavelength range $[4482.6,4483]\AA$. }
\end{table*}

\subsection{Fundamental parameters and abundances}

\begin{figure*}[t!]
\begin{center}
\rotatebox{90}{\resizebox{5.8cm}{!}{\includegraphics{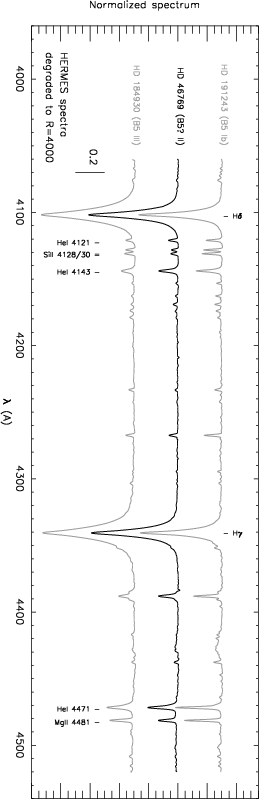}}}
\end{center}
\caption{Optical spectrum of HD\,46769 compared with the spectrum of two
  spectroscopic standards, all brought to a common
  resolution of 4000. The question mark indicates the underabundance of
  \ion{Mg}{}  and
  \ion{Si}{} with respect to the solar values, as discussed in the text.}
\label{SpT}
\end{figure*}

The average HERMES spectrum was used for spectral classification and for the
determination of the fundamental parameters and abundances of the star in a
comparison with the two spectroscopic standards: 
HD\,184930 (B5\,III, Lesh et al.\ 1968) and HD\,191243
(B5\,Ib, Morgan et al.\ 1955). For each of these standards 
we took one HERMES spectrum.
Although Lesh (1968) classified HD\,46769 as B8Ib star, our spectra lead us to
suggest a spectral type of B5II. This is illustrated in Fig.\,\ref{SpT}, 
where we
show one HERMES spectrum of HD\,46769 downgraded to a resolution of
4000, which is often used for spectral classifications, along with the single 
HERMES
spectrum that we obtained for each of the 
two spectroscopic standards. The wings of the Balmer lines, as well
as the \ion{Si}{ii}/\ion{He}{i} 
and \ion{Mg}{ii}/\ion{He}{i} line ratios, lead to B5II, although the
underabundance of \ion{Si}{} and \ion{Mg}{} 
compared to the solar value as discussed below is a
caveat for a clear unique assignment of spectral type (hence the question mark
in Fig.\,\ref{SpT}).

\begin{figure}
\begin{center}
\rotatebox{270}{\resizebox{7cm}{!}{\includegraphics{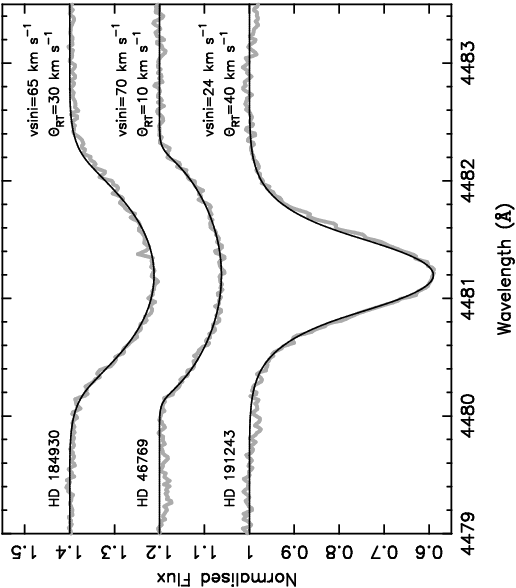}}}
\end{center}
\caption{Zoom in on the \ion{Mg}{ii} 
line at 4481\AA\ of the three stars whose entire
optical spectrum was shown in Fig.\,\protect\ref{SpT}. The values for the
projected rotation velocity $v\sin i$ and for the macroturbulent line broadening
in the approximation of a radial-tangential model ($\Theta_{\rm RT}$) are
indicated in km\,s$^{-1}$. }
\label{MgIIzoom}
\end{figure}

A zoom on the \ion{Mg}{ii} line profile in a spectrum of these three stars
(Fig.\,\ref{MgIIzoom}) reveals that the line profile of HD\,46769 is very
symmetric and mainly broadened by rotation, while the line profiles of
HD\,184930 and HD\,191243 are slightly asymmetric and have broader wings. The
latter phenomenon is often ascribed to the so-called macroturbulent broadening
of hot stars (e.g.\  Howarth et al.\ 1997, Ryans et al.\ 2002, Lefever et
al.\ 2007, Markova \& Puls 2008, Sim\'on-D\'{\i}az et al.\ 2010,
Sim\'on-D\'{\i}az 2011).  
We derived
the rotational broadening, $v\sin i$, and the macroturbulent broadening assuming
an anisotropic radial-tangential model described by $\Theta_{\rm RT}$, following
the definition by Gray (2005), as implemented by Sim\'on-D\'{\i}az \& Herrero
(2007) and further developed by one of us (SSD).  This most recent procedure is
based on a combined Fourier transform and profile fitting analysis, which 
consider
rotational and macroturbulent broadening in addition to the intrinsic line
broadening and the instrumental broadening due to the characteristics of the
spectrograph.  The derived values are indicated in Fig.\,\ref{MgIIzoom} for the
three stars. Ignoring macroturbulent broadening for HD\,46769 still gives a good
fit to the \ion{Mg}{ii} line if the $v\sin i$ is increased by
2\,km\,s$^{-1}$. However, 
it leads to far too narrow wings for the other two stars.

We used the stellar atmosphere code FASTWIND (Santolaya-Rey et al.\ 1997, Puls
et al.\ 2005) to perform a quantitative spectroscopic analysis of HD\,46769 and
the two standards, as well as 
to determine their 
stellar parameters and abundances. The atomic models presently
available for FASTWIND computations are \ion{Si}{}, \ion{Mg}{}, \ion{C}{}, 
\ion{N}{}, and \ion{O}{}.  The stellar
parameter and abundance determinations were done in a two-step process,
following an automatic spectroscopic synthesis strategy similar to the one
described in Castro et al.\ (2012). We used a precomputed grid of FASTWIND
synthetic spectra, which included
lines of \ion{H}, \ion{He}{i-ii}, \ion{C}{ii-iii},
\ion{N}{ii}, \ion{O}{ii-iii}, \ion{Si}{ii-iv}, and \ion{Mg}{ii}. 
More information about the grid
characteristics, coverage, and model atoms incorporated in our
FASTWIND computations can be found in Castro et al.\ (2012).
Table\,\ref{castro1} summarizes the main diagnostic
lines considered here.  The optical spectrum of HD\,46769 includes two
ionization stages for the element \ion{Si}{} 
only, but the \ion{Si}{iii} triplet near
4567\AA\ is very weak. Hence, for the determination of the effective temperature
we relied on the dependencies of the lines of the available ions of this stellar
parameter. In particular, we have found the \ion{He}{i} lines, in combination
with \ion{H}{}, 
to be the best $T_{\rm eff}$ diagnostics due to their strong dependence
on $T_{\rm eff}$ and weak dependence on gravity for mid-B stars.

\begin{table}
\tabcolsep=2pt
\caption{\label{castro1}Diagnostic lines considered for the quantitative
  spectroscopic analysis of HD\,46769.}
\begin{center}\begin{tabular}{ll}
\hline\hline
& Spectral lines (\AA)\\
\hline
\ion{H}{} & H$\alpha$, H$\beta$, H$\gamma$, H$\delta$\\
\ion{He}{i} &  4026, 4387, 4471, 4713, 4922, 5015, 5048, 5875, 6678\\
\ion{Si}{ii} & 3856/62, 4128/30, 5041/56, 6347/71 \\
\ion{Si}{iii} &  4552, 4567, 4574\\
\ion{C}{ii} & 4267, 6578/82\\
\ion{Mg}{ii} &  4481\\
\ion{O}{ii} & 4414/16/52, 4317/19/66, 4641/50/61\\
\ion{N}{ii} & 3995\\
\hline
\end{tabular}
\end{center}
\end{table}

\begin{figure*}[t!]
\begin{center}
\rotatebox{270}{\resizebox{12cm}{!}{\includegraphics{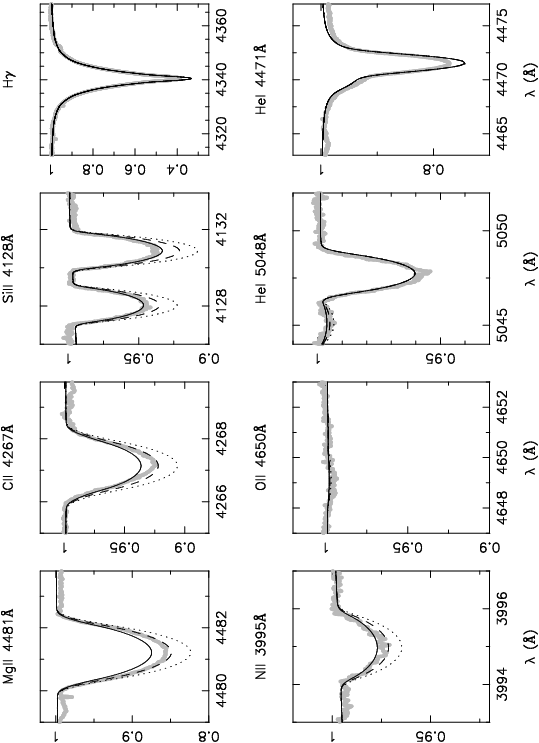}}}
\end{center}
\caption{Comparison of the line predictions for HD\,46769, based on FASTWIND
  atmosphere models for the stellar and abundance parameters listed in
  Table\,\protect\ref{castro2} (dashed). The metal abundances were lowered
  with 0.2 (full) and increased with 0.2 (dotted). In each case, the \ion{He}{}
  abundance was fixed according to n(He)/n(H)=0.1.}
\label{fastwind}
\end{figure*}

In a first step, we derived the stellar parameters of the star, giving more
weight to the \ion{H}{} 
and \ion{He}{i} lines and scaling the abundances of the other
elements to the metallicity of the models. The gravity was determined from the
H$\gamma$ line.  In a second step, we performed the abundance analysis.
The
stellar parameters were fixed to the values derived in step 1, and the
abundances of \ion{C}{}, \ion{N}{}, \ion{O}{}, \ion{Si}{}, and \ion{Mg}{} 
were varied independently to create a set of
synthetic spectra used for comparison with the observed spectra.  In order to
have consistency in the comparison for the three stars, we used only one HERMES
spectrum for each of them.  The results of this analysis are
summarised in Table\,\ref{castro2}. The derived stellar parameters for HD\,46769
are in very good agreement with those determined by Lefever et al.\ (2007). This
is as expected, since we used the same stellar atmosphere code and spectroscopic
analysis techniques but our work is based on more extended and higher quality
spectra. Some line fits for our target star are shown in Fig.\,\ref{fastwind}.

\begin{table*}
\tabcolsep=8pt
\caption{\label{castro2}Derived abundances of HD\,46769, HD\,191243, and
  HD\,184930. }
\begin{center}\begin{tabular}{rccccc}
\hline\hline
Parameter & HD\,46769 &  HD\,191243 &  HD\,184930 & Solar &  B stars in solar
neighbourhood\\ 
\hline
$T_{\rm eff}$ (K) & 13\,000$\pm$1\,000 & 14\,000$\pm$1\,000 & 13\,000$\pm$1\,000  &  --- & --- \\
$\log g$ &  2.7$\pm$0.1  & 2.5$\pm$0.1 & 3.5$\pm$0.1  & --- & --- \\
\ion{Si}{} &      -5.49  &  -4.30  &    -5.10  &   -4.45 & -4.50  \\      
\ion{Mg}{} &      -5.47  &  -4.60  &    -5.10  &   -4.36 & -4.43 \\
\ion{C}{}  &      -4.11  &  -3.30  &    -3.80  &   -4.53 & -3.65 \\      
\ion{N}{}  &      -3.92  &  -3.80  &    ---  &   -4.13 & -4.18      \\
\ion{O}{}  & --- & -3.30 &  --- &  -3.27 &   -3.23 \\
\hline
\end{tabular}
\end{center}
\tablefoot{Metal abundances are listed as $\log (N_{\rm ion}/N_{\rm tot})$ and
  have an uncertainty of 0.2\,dex. In each of the cases, n(He)/n(H)=0.1 was
  kept fixed and the mass loss parameter $\log Q$ (see Kudritzki \& Puls 2000
  for a definition) was below -13.5.  The microturbulent velocity was
  10\,km\,s$^{-1}$ for HD\,46769 and HD\,191243, while it was 5\,km\,s$^{-1}$ for
  HD\,184930.  The solar values were taken from Asplund et al.\ (2009) and those
  of B stars in the solar neighbourhood from Sim\'on-D\'{\i}az (2010), Nieva \&
  Sim\'on-D\'{\i}az (2011), 
and Nieva \& Przybilla (2012). It was not possible to
deduce an \ion{O}{} abundance for HD\,46769 and HD\,184930 because their spectra
do not show \ion{O}{ii} lines, while FASTWIND does not yet have the option to
predict \ion{O}{i} lines.}
\end{table*}

We find an 
underabundance of \ion{Mg}{} and \ion{Si}{} with 0.9\,dex compared to the solar
values (Asplund et al.\ 2009) with $\sim$\,0.2\,dex compared to the \ion{Si}{} 
and \ion{Mg}{}
abundances of B-type stars in the Orion nebula (Sim\'on-D\'{\i}az 2010; Nieva \&
Sim\'on-D\'{\i}az 2011) and, more generally, in the solar neighbourhood (Nieva
\& Przybilla 2012).  
While HD\,191243 has essentially solar abundances,
HD\,184930 is also \ion{Mg}{} and \ion{Si}{} depleted, but less so than
HD\,46769. 
We also
investigated the impact of the uncertainties in $T_{\rm eff}$ and $\log\,g$ on
the derived abundances and found that a variation of 1000\,K and 0.1\,dex in
both parameters leads to a maximium variation of $\sim$\,0.2\,dex in the
\ion{Si}{}  and
\ion{Mg}{} 
abundances. This is well below the variation needed to bring the measured
\ion{Mg}{} 
and \ion{Si}{} 
abundances derived for HD\,46769 in agreement with the solar values.

We come to the conclusion that HD\,46769 is an evolved, moderately rotating
($v\sin\,i\simeq 70$\,km\,s$^{-1}$) B5II star with a normal \ion{He}{} 
abundance but
with an underabundance of 0.9\,dex in \ion{Mg}{} and \ion{Si}{}. 
On the main sequence, such \ion{Si}{}
and \ion{Mg}{} 
deficiencies occur in the so-called HgMn stars (e.g.\  Smith 1993), which
are slow rotators with accompanying strong overabundances in elements such as
\ion{Hg}{}, \ion{Mn}{}, \ion{Cr}{}, \ion{Ti}{}, and \ion{Y}{}. 
They are, however, 
underabundant in \ion{He}{}. These properties are
interpreted in terms of atomic diffusion processes, more particularly the
competition between gravitational settling and radiative levitation. It
remains unsure whether binarity and/or magnetic fields play a role in the
creation and time-evolution of the chemical spots on the surface (e.g.\  Briquet
et al.\ 2007, 2010; Kochukhov et al.\ 2011, Hubrig et al.\ 2012). We checked
explicitly for \ion{Mn}{} and \ion{Hg}{} 
lines in the average spectrum of HD\,46769, but could not detect any.
Other than the HgMn stars, also Bp stars exhibit surface
spots of elements such as \ion{He}{} and \ion{Si}{}, 
giving rise to periodic mmag-amplitude
photometric variations and clear line-profile variability with periodicities of
days (e.g.\  Briquet et al.\ 2004).  Below, we investigate if HD\,46769 is a
line-profile variable.

\subsection{Time series analysis}
\begin{figure}
\rotatebox{270}{\resizebox{4.58cm}{!}{\includegraphics{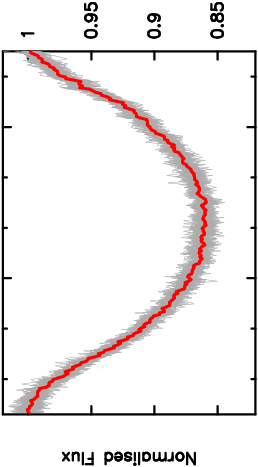}}}
\rotatebox{0}{\resizebox{9cm}{!}{\includegraphics{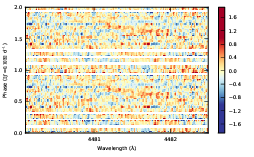}}}
\caption{Top: time series of the \ion{Mg}{ii} 
line at 4481\AA\ distilled from the HARPS
  spectra, with a time base of 25.2\,d. The red thick line is the average
  \ion{Mg}{ii} 
  line. The bottom panel shows the residual \ion{Mg}{ii} 
line profiles with respect to
  the average line profile as a function of the rotational phase and
  represented in a colour scale, 
where we applied a sliding boxcar smoothing with
  a bin width of 0.05\AA\ and a step size of 0.005 cycles; for better visibility
we show two cycles.}
\label{timeseries_MgII}
\end{figure}

\begin{figure}
\begin{center}
\rotatebox{270}{\resizebox{6cm}{!}{\includegraphics{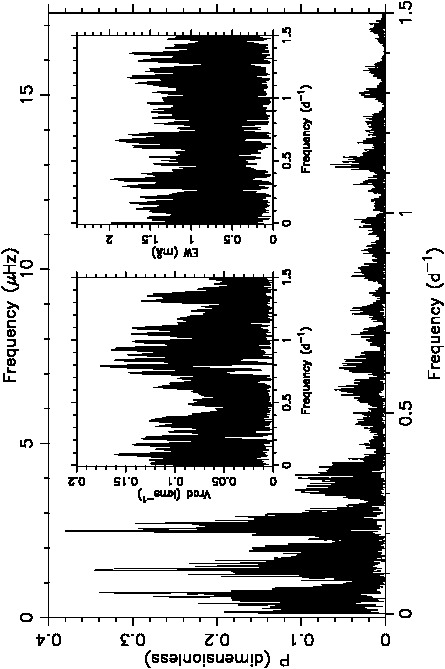}}}
\end{center}
\caption{Product of the Scargle periodograms of the CoRoT and Hipparcos
  light curves and of the EW (right inset) of the
  \ion{Mg}{ii}\,4481\AA\ line, after normalizing those three individual
  periodograms through division by the maximal amplitude (unitless, value
  between 0 and 1). The Scargle periodogram of the radial velocity of the
  \ion{Mg}{ii}\,4481\AA\ line is shown in the left inset for comparison.}
\label{ftproduct}
\end{figure}

In order to search for short-term periodicity in the spectroscopy and compare it
with the detected photometric variability, we first did a visual inspection of
the line profiles of various chemical elements in the observed spectra for the
lines listed in Table\,\ref{castro1}, as well as some additional metal lines.
This did not reveal any obvious variability in the line profile shapes. In order
to quantify the line-profile variations, we performed a line-profile analysis by
means of line moments (e.g.\  Aerts et al.\ 1992 and Briquet \& Aerts 2003) and
the pixel-by-pixel method (Zima 2006). We refer to Aerts
et al.\ (2010a, Chapter 6) for full details of such analyses. An important
aspect of such applications is the selection criterion 
{ to identify} the best
spectral lines, particularly when line variability is not obvious by visual
inspection. In practice, one looks for the least blended lines with considerable
strength and least affected by pressure broadening and effective temperature
variations; see De Ridder et al.\ (2002) and Aerts \& De Cat (2003) for thorough
discussions on these issues for line-profile studies of confirmed pulsating B
stars. For the effective temperature of HD\,46769 and from the gathered
spectroscopy, we identified the \ion{Mg}{ii} doublet near 4481\AA\ as best
suited in terms of S/N level, followed by the \ion{C}{ii} doublet near 4267\AA\ 
and the \ion{Si}{ii} doublet near 4130\AA.  We show the results obtained for the
\ion{Mg}{ii}, which are completely confirmed by the other metal lines.

Fig.\,\ref{timeseries_MgII} shows the time series of the \ion{Mg}{ii} line
deduced from the HARPS spectra, along with the time-average profile of that line
and the residuals after subtraction of the average represented in a colour-scale
plot phased according to the most likely rotation period of the star. It can be
seen that no line patterns stand out. This was also the conclusion from a 2D
frequency analysis with the pixel-by-pixel method done with the software package
{\sc FAMIAS} (Zima 2008).  Subsequently, we resorted to dedicated quantities to
study line-profile variability, according to 
the definition by Aerts et al.\ (1992).  The
time series of the zeroth-order moment (which is the equivalent width of the
line, further abbreviated as EW) and the first moment (indicated as $V_{\rm rad}$
since it is connected with the radial velocity) derived from the \ion{Mg}{ii}
line deduced for all 86 HERMES and HARPS spectra are shown in Fig.\,\ref{lc},
where they were phased with the dominant frequency found in the CoRoT light
curve. The outcome of the Scargle periodograms for the EW and $V_{\rm rad}$ are
shown as insets in Fig.\,\ref{ftproduct}. Both these figures illustrate again
that HD\,46769 does not reveal obvious periodic spectroscopic variability over a
total time base of 1057\,d. Noteworthy is that the maximum frequency peak in
$V_{\rm rad}$ corresponds with $1-4f$ within the frequency resolution, having an
amplitude of 0.15$\pm$0.05\,km\,s$^{-1}$ at 2.3$\sigma$.  We cannot relate any
of the highest peaks in the Scargle periodogram of the EW to the photometric
frequencies.  Enforcing a fit with one of the photometric frequencies leads to
the highest amplitude for $4f$, but the least-squares fit gives an insignificant
amplitude estimate at only 1.9$\sigma$.

Even though the frequency $f$ or its multiples is not significantly present in
the spectroscopy alone, we can still check if its behaviour is consistent with
the CoRoT data or is only connected with noise. We thus applied the same
analysis as already done by Aerts et al.\ (2006) in their search for
low-amplitude variability in the $\beta\,$Cep star $\delta\,$Ceti by combining
MOST and Hipparcos space photometry with ground-based spectroscopy. Since 
we expect flux variations to occur in both the photometry and in the
EW of specral lines, 
we multiplied the EW Scargle periodogram with the one of the CoRoT and of the 
Hipparcos data, after dividing each of those three Scargle periodograms by their
maximum, i.e.\  we computed
$$\displaystyle{P (f) = 
\frac{P_{\rm Scargle}^{\rm EW}(f)}{P_{\rm Scargle}^{\rm EW}(f_{\rm max}^{\rm EW})}
\cdot 
\frac{P_{\rm Scargle}^{\rm Hipparcos}(f)}
{P_{\rm Scargle}^{\rm Hipparcos}(f_{\rm max}^{\rm Hipparcos})}
\cdot 
\frac{P_{\rm Scargle}^{\rm CoRoT}(f)}
{P_{\rm Scargle}^{\rm CoRoT}(f_{\rm max}^{\rm CoRoT})}.}
$$ In this way, we give equal weight to the CoRoT, Hipparcos, and HERMES+HARPS
data in the search for consistent frequencies in those three completely
independent data sets.  The idea is that all the peaks due to noise will keep on
producing a noisy periodogram after multiplication of the normalized
periodograms. In contrast, 
peaks that are consistently present in the data, even
though not significantly above the noise level in the individual Scargle
periodograms, will still keep a value that is clearly above the noise level
when full compatibility between the three Scargle periodograms occurs. Prior to
normalization, the Scargle periodograms were computed with a frequency step
according to the best Rayleigh limit, which occurs for the Hipparcos
data set. The outcome of this product of the three normalized Scargle
periodograms, which is a dimensionless quantity with a value between zero (bad
frequency) and one (frequency dominant in all three independent data sets), is
shown in the main panel of Fig.\,\ref{ftproduct}. It can be seen that three sets
of peaks stand out, i.e.\  that the frequencies deduced from the CoRoT light
curve ``survive'' this consistency test. These occur at 0.2063, 0.1083, and
0.0522\,d$^{-1}$, where the first one coincides with $4f$ and all three are
equal to the frequencies in the CoRoT data listed in Table\,\ref{sn} within the
frequency resolution.  Hence, the spectroscopy and Hipparcos data by themselves
do not reveal significant frequencies but are consistent with the low-amplitude
periodic variability detected in the CoRoT light curve.

\subsection{Spectropolarimetric measurements}

Given the photometric variability and the abundance pattern of HD\,46769, as
well as the debate about the role of magnetic fields in chemically anomalous
rotationally variable main-sequence stars, we made use of the Narval
spectropolarimeter installed at the 2\,m TBL telescope at 
the Pic du Midi in France to
search for the presence of a magnetic field in HD\,46769. On December 2012 we
obtained two Stokes V spectra computed by constructively combining four
subexposures of 900 seconds, each taken in different configurations of the
polarimeter wave plates. The data reduction was performed using {\sc
  libre-esprit} (\'Echelle Spectra Reduction: an Interactive Tool;
Donati et al.\ 1997).

For each of the two measurements, we applied the least-squares deconvolution
(LSD) technique (Donati et al.\ 1997) to the photospheric spectral lines in the
whole \'echelle spectrum ($\lambda\in [3750,10500]\,$\AA) in order to construct
single averaged Stokes I and V profiles with an increased S/N ratio. To this
end, we used a mask including photospheric He and metal lines of various
chemical elements, which we created from the Kurucz atomic database and ATLAS~9
atmospheric models of solar abundance (Kurucz 1993) for T$_{\rm
  eff}$=13000 K and $\log g$=3.0, with intrinsic line depths larger than
0.1. Two adjustments were applied to the original mask. First, we removed lines
from the mask that are not suited for the LSD analysis, i.e.\ 
those that are blended
with the Balmer lines or with atmospheric telluric lines and weak lines
that are
indistinguishable from the noise. Second, the relative depths of all spectral
lines used in the mask were modified to correspond to those of the
observed spectral lines. The final mask contains 311 lines, 
and the two resulting
LSD Stokes V profiles have S/N values of 5864 and 5309 per 2.6\,km\,s$^{-1}$
pixels. A visual inspection of each of them shows no signature of a magnetic
field.

We used the LSD Stokes I and V profiles to compute the line-of-sight component
of the magnetic field integrated over the visible stellar surface, i.e.\  the
longitudinal magnetic field in Gauss, given e.g.\ by Eq.\,(1) of Wade et
al.\ (2000). In this equation we used a mean wavelength $\lambda$=500\,nm and a
mean Land\'e factor $g$=1.13. For the integration limits, a range of 70
km\,s$^{-1}$ around the line centre was adopted. We obtained longitudinal field
values of $B_{l_1}$=-9.4$\pm$21.0\,G and $B_{l_2}$=22.1$\pm$22.9\,G at
Mid-HJD$_1$=2456271.657354 and Mid-HJD$_2$=2456272.627450, respectively, i.e.\ 
the longitudinal field values are consistent with the absence of a magnetic
field.

As we only have two measurements available, we cannot definitely exclude the
presence of a field in the star. Indeed, it is possible that a magnetic field
exists with a strength below our detection level. Therefore we performed a Monte
Carlo analysis of the spectropolarimetric data to derive the upper limit of the
polar magnetic field that could be present at the surface of HD\,46769.

To perform our simulation we assumed a centred oblique dipole field model, for
which the free parameters are the stellar inclination angle $i$ (in degrees),
the magnetic obliquity angle $\beta$ (in degrees), the dipole magnetic intensity
$B_{\rm pol}$ (in Gauss), and a phase shift $\Delta\Phi$ compared to the adopted
stellar ephemeris. In our model, we use Gaussian local intensity profiles with a
width calculated using the resolving power of Narval and a line-broadening value
of 20 km\,s$^{-1}$, representing the overall effect of the intrinsic line widths
as well as micro- and macroturbulence.  The depth is determined by fitting the
observed LSD Stokes I profiles. The local Stokes V profiles are calculated
assuming the weak-field case and using the weighted mean Land\'e factor and
wavelength of the lines derived from the LSD mask (see above). The synthetic
Stokes V profiles are obtained by integrating over the visible stellar surface
by using a linear limb-darkening law with a parameter equal to 0.386
(Claret \& Bloemen 2011). The Stokes V profiles are normalised to the continuum
intensity.

\begin{figure}[!ht]
\begin{center}
\resizebox{\hsize}{!}{\includegraphics[clip]{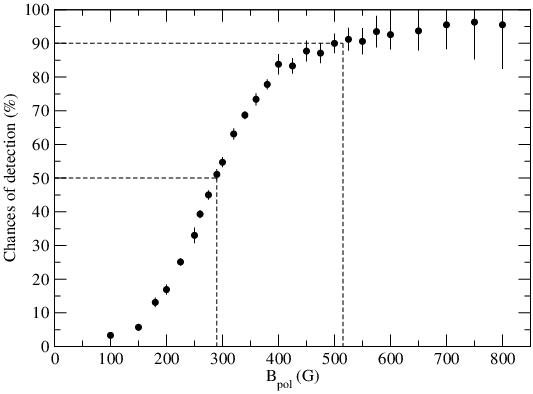}}
\caption[]{Chances of detection of a magnetic field in at least one of the two
  Narval measurements of HD\,46769, depending on the polar field strength,
assuming $P_{\rm rot}=$ 9.69\,d.}
\label{maglimit}
\end{center}
\end{figure}

We considered the two possible values for the rotation period $P_{\rm rot}=$ 4.84
or 9.69\,d and various values for the polar field strength. We then
simulated 1000 dipole models with random values of $i$, $\beta$, 
and $\Delta\Phi$
for each $B_{\rm pol}$ value. We simulated the two Stokes V profiles
simultaneously with a noise corresponding to the S/N ratio of the Narval
data. We then computed among each set of 1000 models the chances of detecting
the field in at least one of the two measurements. We find that there was a 50\%
(90\%) chance of detection at 3$\sigma$ of a field with any oblique dipolar
configuration and a polar strength above $\sim$290 G ($\sim$515 G) for the case
of $P_{\rm rot}=$ 9.69\,d. This is illustrated in 
Fig.\,\ref{maglimit}, where we show the probability
of detection of a field according to its strength in at least one of the two
available measurements of HD\,46769, assuming $P_{\rm rot}=$ 9.69\,d.  Results
are similar, whichever of the two possible rotation periods we consider.

We conclude that an oblique magnetic dipole field of a few hundred Gauss at the
poles { could be} present while it remained undetected in the two Narval
measurements. A stronger magnetic field can be excluded, unless the hypothetical
field were more complex than dipolar. However, for massive non-Bp stars,
detected fields are usually observed to be dipolar (e.g.\ Neiner et al.\ 2003)
and also theoretically predicted to be dipolar (Duez \& Mathis 2010).
Non-dipolar examples (e.g.\ Kochukhov et al.\ 2011) remain very rare so far.
Our results are consistent with the scarcity of magnetic fields in OBA
supergiants, despite intensive efforts to detect them (e.g.\ Verdugo et
al.\ 2003, Hubrig et al.\ 2013, Shultz et al.\ 2013).

\section{Discussion}

All the data taken together lead us to conclude that HD\,46769 is a
low-amplitude, rotationally modulated, photometrically variable B5II
supergiant. The photometric variability of some 100\,ppm level is interpreted in
terms of low-contrast chemical and/or temperature inhomogeneities. This gives
 rise to
rotational modulation in the light curve, which remains undetectable in
high-precision spectroscopy alone.  

{ The fundamental parameters of the star place it just outside pulsational
  instability regions in the $(T_{\rm eff},\log\,g)$ diagram (e.g.\  Fig.\,10 of
  Lefever et al.\ 2010), although the error bars cover the red edge of excited
  gravity modes computed by Saio et al.\ (2006).}
Despite the
absence of pulsations with a detection threshold of some 10\,ppm in photometric
precision, the star is an interesting object for its underabundance in Mg and
Si, and also in the context of the occurrence of macroturbulence in evolved
massive stars whose physical cause remains unknown.  Indeed, HD\,46769 is a
moderate rotator, 
and its spectral lines have very sharp wings in comparison with
more luminous stars of similar temperature. We found an anisotropic
radial-tangential macroturbulence of less than 
10\,km\,s$^{-1}$, i.e.\  far lower than
the projected rotational velocity broadening. A good fit to the line wings would
also be achieved without considering macroturbulent broadening, increasing the
$v\sin\,i$ by only a few km\,s$^{-1}$. In contrast, to be able to 
explain their spectral line wings,
the majority of the evolved
variable B supergiants require large macroturbulent velocities, several of which
are supersonic and above the rotational velocity. 
Aerts et al.\ (2009) suggested that the
detected high level of macroturbulence needed to explain the line wings of
massive supergiants in high-precision spectroscopy could be due to the
collective effect of numerous low-amplitude gravity-mode pulsations, which are
unresolved individually and give rise to a global line broadening
in single-snapshot spectra.  This idea
is currently being investigated in more detail from a systematic long-term
observational study of a large sample of OB-type stars from high-precision
time-series spectroscopy (e.g.\  Sim\'on-D\'{\i}az et al.\ 2010, 2012 and its
follow-up work, in preparation).  Our results for the case of HD\,46769 are
consistent with this interpretation in the sense that it concerns an evolved
non-pulsating B-type supergiant with sharp line wings, for which macroturbulence
does not need to be invoked to get a good fit of the spectral line wings.

\begin{acknowledgements}
The authors thank the MiMeS collaboration for having provided
Narval observations.
  The research leading to these results received funding from the ERC under the
  European Community's 7th Framework Programme (FP7/2007--2013)/ERC grant
  agreement n$^\circ$227224 (PROSPERITY), as well as from the Belgian Federal
  Science Policy Office BELSPO (C90309: CoRoT Data Exploitation).  
SS-D thanks I.\ Negueruela for an interesting discussion about the spectral
classification of HD\,46769 and acknowledges financial support from the Spanish
Ministry of Economy and Competitiveness (MINECO) under the grants
AYA2010-21697-C05-04, Consolider-Ingenio 2010 CSD2006-00070, Severo Ochoa
SEV-2011-0187, and by the Canary Islands Government under grant PID2010119.
E.P.\ and
  M.S.\ acknowledge financial support from the Italian PRIN-INAF 2010 {\it
    Asteroseismology: looking inside the stars with space- and ground-based
    observations.}  M.R.\ acknowledges financial support from the FP7 project
  {\it SPACEINN: Exploitation of Space Data for Innovative Helio- and
    Asteroseismology}.
\end{acknowledgements}



\end{document}